\documentstyle[12pt,psfig]{l-aa}
\def\msun{M_{\odot}}

\def \rsun {\ifmmode$R$_{\odot}\else R$_{\odot}$\fi}

\def \hcm {\hbox {\ifmmode $ H atoms cm$^{-2}\else H atoms cm$^{-
2}$\fi}}
\def \src {4U\,1626$-$67}
\def\approxgt{\mathrel{\hbox{\rlap{\lower.55ex \hbox {$\sim$}}
        \kern-.3em \raise.4ex \hbox{$>$}}}}
\def\approxlt{\mathrel{\hbox{\rlap{\lower.55ex \hbox {$\sim$}}
        \kern-.3em \raise.4ex \hbox{$<$}}}}

\newcommand {\sax} {{\it BeppoSAX}}

\newcommand {\PdotP} {${\rm \dot{P}/P}$}

\begin{document}

\thesaurus{ (08.02.1; 08.09.2; 08.14.1; 13.25.5)}

\title{The complex X-ray spectrum of the low-mass X-ray binary \src}

\author{Alan Owens \and T. Oosterbroek \and A.N. Parmar}

\institute{Astrophysics Division, Space Science Department of ESA, 
ESTEC, P.O. Box 299, 2200 AG Noordwijk, Netherlands}
  
\date{Submitted on:}
\offprints{A. Owens: aowens@astro.estec.esa.nl}

\maketitle
\markboth{A. Owens et al.: X-ray observations of \src}{A. Owens et al.: 
X-ray observations of \src}

\begin{abstract}
We report on observations of the X-ray pulsar \src\ by the LECS 
instrument on-board \sax. We confirm the recent ASCA discovery
of excess emission near 1~keV (Angelini et al. 1995). The 
pulse period of 7.66794 $\pm$ 0.00004~s indicates that the source 
continues to spin-down. The phase averaged spectrum is well fit by 
an absorbed power-law of photon index 0.61 $\pm$ 0.02 and a blackbody 
of temperature 0.33 $\pm$ 0.02~keV, together with an emission 
feature at 1.05 $\pm$ 0.02~keV. This spectral shape is similar to 
that observed by ASCA during the spin-down phase, but significantly 
different from measurements during spin-up. This 
suggests that the change in spectrum observed by ASCA may be a stable 
feature during spin-down intervals. The source intensity is a factor 
$\sim$2 lower than observed by ASCA three years earlier, confirming 
that \src\ continues to become fainter with time.  
\end{abstract}

 
\section{Introduction}

The X-ray source \src\ is a 7.7 s pulsar in a highly compact binary 
system of orbital period 2485~s. It is unusual in that it is one of the 
few low mass X-ray binary systems to contain an X-ray pulsar.
While the X-ray emission is strongly modulated by the pulsar, there is 
no evidence for Doppler shifts induced by the orbital motion of the 
source, despite extensive searches. The implies that the projected orbital 
radius of the neutron star is small, i.e., $a_x$sin $i$ $\le$ 13 
m-lt-s (Levine et al. 1988). Optical pulsations were first detected 
by Ilovaisky et al. (1978) and interpreted as X-ray 
re-processing near to, or along, the line of sight to the X-ray source. 
Middleditch et al. (1981) found a single low frequency side-lobe 
which they interpret as arising from the optical re-processing of the 
primary X-rays on the companion star. Assuming the 
pulsar spins in the same sense as the orbital motion, these photons will 
be shifted to a lower frequency by the rotation frequency of
the binary orbit. From the observed frequency shift of 0.4~mHz 
an orbital period of 2485~s and a projected semimajor axis 
of 0.4 lt-s is inferred. The current picture of \src\ is of a highly 
compact system comprising a neutron star of mass $\sim$1 M$_\odot$, with
a 0.08$\msun$ Main Sequence or 0.02$\msun$ white dwarf
companion (Verbunt et al. 1990).
 
For the first decade after its discovery \src\ was rapidly spinning-up
at a rate of \PdotP $\sim-2 \times 10^{-4}$~yr$^{-1}$. However,
long term monitoring by the Burst and Transient Source Experiment
(BATSE) on-board the Compton Gamma-ray Observatory beginning in 1991
April found that ${\rm \dot{P}}$, and hence the accretion torque,
had changed sign, resulting in a spin-down at nearly the same rate (Wilson
et al. 1993). It is estimated that the reversal occurred in mid-1990.
Observations of \src\ during the earlier spin-up phase found that the 
phase averaged spectrum could be modeled by a blackbody of temperature, 
kT, $\sim$0.6 keV together with a power-law of photon index, $\alpha$, of 
$\sim$1 (e.g., Pravdo et al. 1979; Kii et al. 1986). In the 2--10~keV 
energy range the pulse profile consisted of a narrow pulse with a ``notch'',
while at higher and lower energies this evolved into a roughly 
sinusoidal shape (Levine et al. 1988; Mavromatakis 1994). 
This strong energy dependence may result from anisotropic radiative 
transfer in a strongly magnetized plasma (Kii et al. 1986). 
 
In addition to periodic pulsations, \src\ also exhibits quasi periodic 
behavior. Both the X-ray and optical intensities show correlated flaring 
on timescales of $\sim$1000~s (Joss et al. 1978). The origin of this 
behavior is unknown. A 40 mHz quasi periodic oscillation (QPO) has been 
detected in X-rays (Shinoda et al. 1990) and more recently in the optical 
band (Chakrabarty et al. 1997). 

Finally, the recent observation of an emission line complex near 1.0 keV by 
Angelini et al. (1995) is particularly interesting. This emission is 
interpreted as arising primarily from Ne K rather than from Fe L, 
based on the measured line energies and intensities. This result suggests 
that the companion star has evolved past its hydrogen burning stage. Ne 
is a by-product of He burning and therefore its overabundance suggests 
that the star is burning, or has burnt, He.
  
\section{Observations}
 
The Low-Energy Concentrator Spectrometer (LECS) is one of 5 scientific 
instruments on-board the \sax\ satellite (Boella et al. 1997). 
The LECS is described in Parmar et al. (1997) and consists of 
a nest of conical approximation to Wolter I X-ray mirrors
which produce a focused beam of X-rays on an imaging gas scintillation 
proportional counter. The detector employs a driftless design which, 
in conjunction with an extremely thin entrance window, ensures an extended 
low energy response. The nominal energy range of the instrument is 0.1--10~keV 
and the full width at half maximum (fwhm) energy resolution is 19\% at 1 keV. 
The field of view (fov) is 37$'$ diameter and the spatial resolution is 
5.1$'$ fwhm at 1~keV. \sax\ was launched into a 600 km equatorial orbit 
on 1996 April 30.
 
\src\ was observed twice by the LECS during the Science
Verification Phase on 1996 August 6 from 12:00 UT to 21:50 UT 
and on August 9 02:25 UT to August 11 00:00 UT. The total source 
exposure is 40 ks and the mean LECS count rate is $1.029 
\pm 0.005$~s$^{-1}$. Because of the large spatial 
extent of the point spread function, background data 
were acquired separately by viewing blank regions of 
sky. Background subtraction is not critical for such a bright source.

\label{4uspec}

\begin{figure}
\centerline{\psfig{figure=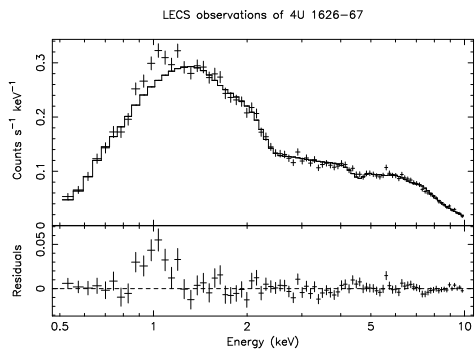,width=7cm,angle=-90}}
\caption{\protect \small Phase averaged data and folded model for 
\src. The data are described by a model consisting of a blackbody 
Of kT = 0.33 keV, a power-law of $\alpha$ = 0.61, an 
emission-line centered on 1.05~keV and low energy absorption. For 
illustrative purposes, the line intensity is set to zero. The lower 
panel shows the residuals} 
\label{4uspec}
\end{figure}

\section{Spectral analysis}

Source events were obtained from both observations using the standard 
extraction radius of 8$'$, centered on the source centroid. The 
extracted data were rebinned to have bin widths of 0.25 $\times$ 
the fwhm energy resolution. This is to help ensure an unbiased fit across 
the energy range, while preserving sensitivity to line features.  
Since the LECS response is dependent on both position within the
fov and extraction radius, the appropriate matrix was created using
SAXLEDAS version 1.4.0 (Lammers 1997).   

The spectrum was first fit with an absorbed blackbody plus power-law model 
yielding a $\chi^2$ of 115 for 84 degrees of freedom (dof). Inspection 
of the residuals (see Fig. \ref{4uspec}) shows an excess centered around 
1 keV, similar to that seen by Angelini et al. (1995) with the ASCA 
solid-state imaging spectrometer (SIS). The addition of a Gaussian line 
feature to the model results in a $\chi^2$ of 92 for 81 dof.  An 
F-test shows that the improvement is significant at the $\ge$ 99.99\% 
level. The best-fit line energy and flux are 1.05 $\pm ^{~0.01}_{~0.03}$~keV 
and (4.58 $\pm_{~1.2}^{~1.5}$) $\times~10^{-4}$~photons~cm$^{-2}$ s$^{-1}$, 
respectively. (All uncertainties are quoted at 68\% confidence.) 
The equivalent width (EW) is 47.6 $\pm$ 13.5 eV. The spectrum is shown in 
Fig. \ref{4uspec} with the line intensity set to zero. The best-fit values 
of kT and $\alpha$ are 0.33 $\pm$ 0.02~keV and 0.61 $\pm$ 0.02, respectively. 
The derived absorption column is (6.9 $\pm$ 2.0)$~\times~10^{20}$ atoms 
cm$^{-2}$, consistent with the interstellar value in the direction 
of \src\ (Daltabuit \& Meyer 1972; Dickey \& Lockman 1990).

The best-fit line energy agrees well with that expected from a blend 
of 80\% Ne Ly-$\alpha$ (1.021 keV) and 20\% Ne He-$\beta$ (1.084 keV) 
emission, consistent with the ASCA result of Angelini et al. (1995). 
The values of kT and $\alpha$ are also similar to those measured using ASCA, 
but both a factor of 2 lower than earlier HEAO-1 (Pravdo et al. 1979) 
and $Tenma$ (Kii et al. 1986) values. The blackbody radius, r$_b$, of 
1.0 d$_{kpc}$ km, where d$_{\rm {kpc}}$ is the distance in kpc, is also 
consistent with ASCA. The 0.5--10~keV luminosity is 2.0 $\times$ 
10$^{34}$~ergs~s$^{-1}$~d$_{\rm {kpc}}^{2}$ which is $\sim$40\% lower 
than measured by ASCA three years earlier, and a factor of 6 lower than 
that derived using the {\it Einstein} Solid State Spectrometer in 1979 
March (Angelini et al. 1994). In the case of ASCA, Angelini et al. (1995) 
attribute the spectral and luminosity differences to the torque reversal, 
since the HEAO-1, $Tenma$, and {\it Einstein} measurements were carried 
out prior to mid-1990. The ASCA and \sax\ measurements 
were performed during the present spin-down phase.

\begin{table}[htb]
\caption{\protect \small Best-fit spectral 
parameters.  Uncertainties are given at the 68\% confidence level. Line 
energies, widths and temperatures are in units of keV. Line fluxes and 
column densities are in units of 10$^{-4}$ photons cm$^{-2}$ s$^{-1}$ and 
10$^{21}$ atoms cm$^{-2}$, respectively, EW are given in units of eV, and 
r$_b$ has units of d$_{kpc}$ km.} 
\label{lfit}
 
\begin{tabular}{ll} \hline
Parameter &  Value    \\
\multicolumn{2}{c}{Model 1: Absorbed power-law plus blackbody} \\ \hline

N$_H$                                       &  1.10 $\pm$ 0.20\\
$\alpha$                                    &  0.64 $\pm$ 0.02     \\
kT                                          &  0.29 $\pm$ 0.01    \\
r$_b$                                       &  1.4 $\pm$ 0.8    \\
$\chi^2$/$\nu$                              &  115/84              \\ 
\hline
\multicolumn{2}{c}{Model 2: Absorbed power-law, blackbody and line} \\ \hline
N$_H$                                       & 0.69 $\pm$ 0.20     \\
$\alpha$                                    & 0.61 $\pm$ 0.02      \\
kT 					    & 0.33 $\pm$ 0.02   \\
r$_b$  				            & 1.0 $\pm$ 0.5    \\
E$_{line}$                                  & 1.05 $\pm_{~0.03}^{~0.01}$    
  \\
$\sigma_{line}$                             & 0.04 $\pm$ 0.04             
        \\
EW                                          & 48 $\pm$ 14   \\
Flux$_{line}$                               & 4.6  $\pm_{~1.2}^{~1.5}$      \\
$\chi^2$/$\nu$                              & 92/81                \\
\hline
\multicolumn{2}{c}{Model 3: Absorbed power-law, blackbody and lines}  \\
\hline
N$_H$                                       & 0.81 $\pm$ 0.26                  
 \\
$\alpha$                                    & 0.62 $\pm$ 0.02              \\
kT                                          & 0.33 $\pm$ 0.02            
             \\
r$_b$                                       & 1.0 $\pm$ 0.6    \\
O He-$\alpha$  0.568 keV flux               & 5.3  $\pm_{~5.3}^{~8.2}$ \\
EW                                          & 34  $\pm$ 34            \\
Ne Ly-$\alpha$ 1.021 keV flux               & 3.2 $\pm$ 1.5  \\
EW                                          & 28 $\pm$ 12           \\
Ne He-$\beta$  1.084 keV flux               & 1.5 $\pm$ 1.3  \\
EW                                          & 13  $\pm$ 10            \\
Fe K-$\alpha$  6.400 keV flux               & 0.7 $\pm$ 0.5  \\
EW                                          & 38.6 $\pm$ 22.9           \\
$\chi^2$/$\nu$                              & 88/80                         \\
\hline
\end{tabular}
\end{table}

In order to investigate whether the single narrow feature seen by the
LECS is consistent with the emission line complex observed by the
ASCA SIS, the feature at 1.05~keV was replaced by the blend of lines 
used in the ASCA fit (see Table~1 of Angelini et al. 1995).  Each 
line was added separately with its energy and width fixed at the 
measured ASCA values. The only $proviso$ for inclusion in the 
final fit was that $\chi^2$ must reduce. Only three lines satisfied 
this criteria; the Ne Ly-$\alpha$ line at 1.021 keV, the Ne He-$\beta$ 
line at 1.084 keV, and the O He-$\alpha$ line at 0.568 keV. Surprisingly, 
the fit did not require an Ne He-$\alpha$ line and we derive an upper flux
limit at the 90\% confidence level of 2.2 $\times$ 10$^{-4}$ photons 
cm$^{-2}$ s$^{-1}$. Phase-resolved ASCA spectra reveal the presence of 
a weak iron K feature over a narrow range of pulse phases (Angelini et 
al. 1995). Including such a line in the current model gives a $\chi^2$ 
of 88 for 80 dof. Best-fit spectral parameters are listed in Table 1. 
The derived O He-$\alpha$ and Ne He-$\beta$ line intensities are 
consistent with the ASCA values reported in Angelini et al. (1995). 
However, the Ne Ly-$\alpha$ line intensity is a factor of 3 lower 
and the 90\% confidence upper flux limit to any Ne He-$\alpha$ emission 
is lower by a factor of $\sim$ 6.

\label{4u10per}

\begin{figure}
\centerline{\psfig{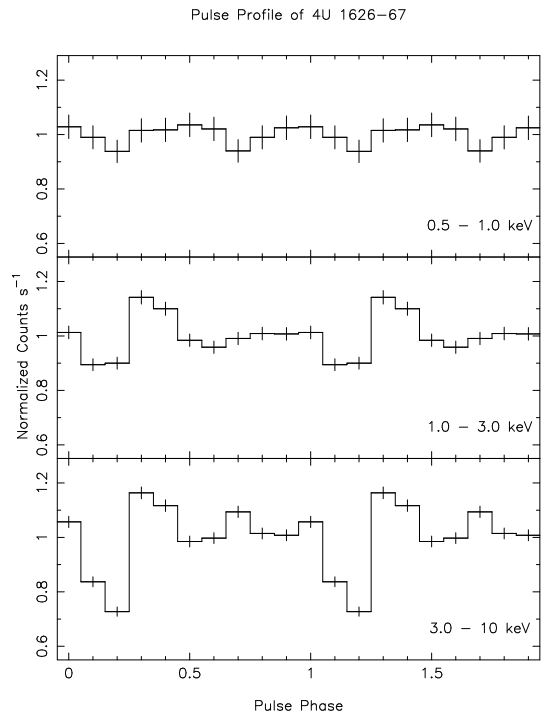}}
\caption{\protect \small LECS Pulse profiles for \src\ folded over the 
best-fit period of 7.66794~s in three energy bands}
 \label{4u10per}
\end{figure}

We next investigated an alternate spectral model for \src\ in which 
the power-law 
plus blackbody is supplemented by emission from an optically-thin 
collisionally ionized plasma (specifically the VMEKAL model in XSPEC 
v.9.01). In principle, this would allow us to estimate the elemental 
abundances necessary to produce the excess emission around 1 keV, in 
a similar manner to Angelini et al. (1995). The abundances of Ne and 
Fe were allowed to vary while the abundances of the other elements 
were fixed at the photospheric values of Anders \& Grevesse (1989). 
Both high Ne and Fe over-abundances gave acceptable fits to the 
data with $\chi^2$'s comparable to the "Ne complex" fit . However, 
the LECS spectrum is of insufficient quality to determine meaningful 
limits to these abundances.

\section{Temporal analysis}
 
The barycentric pulse period during the LECS observations of $7.66794 
\pm 0.00004$~s is in good agreement with the predicted value of 
7.667943 $\pm$ 0.000006~s derived from BATSE data (Chakrabarty 1997). 
The pulse profiles in the energy ranges 0.5--1, 1.0--3.0, and 3.0--10~keV, 
are shown in Fig. \ref{4u10per}.  The overall shape and energy dependence 
are similar to those seen by ASCA. The pulse profile in the lowest energy 
band in Fig. \ref{4u10per} may be consistent with that seen during spin-up 
(e.g., Pravdo et al. 1979), but with a reduced amplitude. 

\label{4ups} 
\begin{figure}
\centerline{\psfig{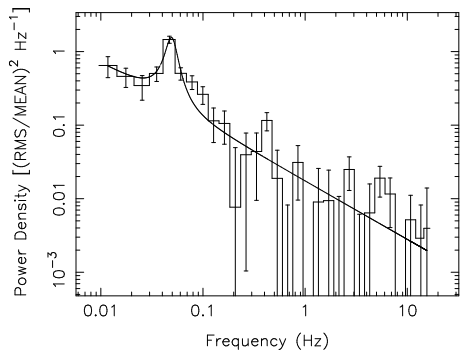}}
\caption{\protect \small The power density spectrum for \src\ in the 
0.1--10~keV 
energy band. A QPO is visible at $\sim$ 0.05 Hz}
\label{4ups}
\end{figure}

Figure~\ref{4ups} shows a white noise subtracted, power density spectrum of 
the LECS data from which quasi periodic oscillations (QPOs) are apparent. 
The center frequency is 0.049 $\pm$ 0.002~Hz and the fwhm is 0.015 
$\pm ^{~0.003} _{~0.006}$~Hz. The QPO strength is 18 $\pm ^{~4}
_{~9}$\% {\it rms} of the mean count rate. Within uncertainties, the QPO 
amplitude is the same in the 0.5--2.0~keV and 2.0--10.0~keV energy ranges. 
The QPO centroid frequency, width, and amplitude are consistent with the ASCA 
measurements (Angelini et al. 1995). The amplitude and width are also 
consistent with the $Ginga$ values (Shinoda et al. 1990), but the centroid 
frequency is not. This change may be related to the torque reversal.

\section{Discussion}

The LECS data confirm the recent ASCA detection of excess emission
near 1~keV from \src. In addition, the best-fit spectral parameters
confirm that the spectral shape remains changed following the mid-1990 
torque reversal. Both kT and $\alpha$ decreased by a factor of $\sim$2, 
while the X-ray luminosity decreased by a factor of 6. As before the 
reversal, the new parameters appear stable with time. Chakrabarty et 
al. (1997) report that the intensity of the source is steadily 
decreasing with time. The LECS results support this, since the 
0.5--10~keV source intensity is 0.6 that measured by ASCA.
Extrapolating from previous measurements (see Fig. 8 of Chakrabarty 
et al. 1997), the expected decrease is a factor of $\sim$0.7. 
The shape of the pulse profile is also different from that measured 
before the reversal. Prior to the reversal, the profile was strongly 
energy and phase dependent (e.g., see Levine et al. 1988), while 
the LECS pulse profile has the same shape, but a variable amplitude, 
over the 1.0--10.0~keV energy range (see Fig. 2).

\begin{acknowledgements}

We thank Chris Butler, Luigi Piro, and the staff of the \sax\ Science Data 
Center in Rome. Lorella Angelini is thanked for discussions and 
Deepto Chakrabarty for providing the BATSE pulse period. T. Oosterbroek 
acknowledges an ESA Research Fellowship. The \sax\ satellite is a joint 
Italian and Dutch programme.

\end{acknowledgements}

\end{document}